\documentclass{article}

\usepackage{arxiv}
\usepackage[utf8]{inputenc} 
\usepackage[T1]{fontenc}    
\usepackage{hyperref}       
\usepackage{url}            
\usepackage{booktabs}       
\usepackage{amsmath}
\usepackage{amsfonts}       
\usepackage{nicefrac}       
\usepackage{microtype}      
\usepackage{lipsum}		
\usepackage{graphicx}
\usepackage{natbib}
\usepackage{doi}
\usepackage{float}          
\usepackage{placeins}       
\usepackage{tabularx}
\usepackage{ragged2e} 
\newcolumntype{Y}{>{\RaggedRight\arraybackslash}X}
\usepackage{subcaption} 

\title{Deciphering Scientific Collaboration in Biomedical LLM Research: Dynamics, Institutional Participation, and Resource Disparities}

\author{
\large \textbf{Lingyao Li}\textsuperscript{1}\thanks{Equal contribution.}
\and \large \textbf{Zhijie Duan}\textsuperscript{2}\footnotemark[1]
\and \large \textbf{Xuexin Li}\textsuperscript{3}
\and \large \textbf{Xiaoran Xu}\textsuperscript{1}
\and \large \textbf{Zhaoqian Xue}\textsuperscript{2}
\and 
\\[0.3em]
\large \textbf{Siyuan Ma}\textsuperscript{3}\thanks{Corresponding author. 2525 West End Avenue, Nashville, TN 37203. \texttt{siyuan.ma@vumc.org}}
\qquad
\large \textbf{Jin Jin}\textsuperscript{2}\thanks{Corresponding author. 423 Guardian Drive, Philadelphia, PA 19104. \texttt{jin.jin@pennmedicine.upenn.edu}}
\\[1.2em]
\textsuperscript{1}School of Information, University of South Florida, Tampa, FL\\
\textsuperscript{2}Department of Biostatistics, Epidemiology and Informatics, University of Pennsylvania, Philadelphia, PA\\
\textsuperscript{3}Department of Biostatistics, Vanderbilt University Medical Center, Nashville, TN
}
\date{}

\renewcommand{\shorttitle}{Deciphering Scientific Collaboration in Biomedical LLM Research}

\hypersetup{
pdftitle={A template for the arxiv style},
pdfsubject={q-bio.NC, q-bio.QM},
pdfauthor={David S.~Hippocampus, Elias D.~Striatum},
pdfkeywords={First keyword, Second keyword, More},
}

\begin{document}
\maketitle

\begin{abstract}
Large language models (LLMs) are increasingly transforming biomedical discovery and clinical innovation, yet their impact extends far beyond algorithmic revolution—LLMs are restructuring how scientific collaboration occurs, who participates, and how resources shape innovation. Despite this profound transformation, how this rapid technological shift is reshaping the structure and equity of scientific collaboration in biomedical LLM research remains largely unknown. By analyzing 5,674 LLM-related biomedical publications from PubMed, we examine how collaboration diversity evolves over time, identify institutions and disciplines that anchor and bridge collaboration networks, and assess how resource disparities underpin research performance. We find that collaboration diversity has grown steadily, with a decreasing share of Computer Science and Artificial Intelligence authors, suggesting that LLMs are lowering technical barriers for biomedical investigators. Network analysis reveals central institutions, including Stanford University and Harvard Medical School, and bridging disciplines such as Medicine and Computer Science that anchor collaborations in this field. Furthermore, biomedical research resources are strongly linked to research performance, with high-performing resource-constrained institutions exhibiting larger collaboration volume with the top 1\% most connected institutions in the network. Together, these findings reveal a complex landscape, where democratizing trends coexist with a persistent, resource-driven hierarchy, highlighting the critical role of strategic collaboration in this evolving field.
\end{abstract}


\section{Introduction}

Since the release of BERT and ChatGPT, large language models (LLMs) have emerged as transformative tools within the biomedical domain \citep{Lan2025,Tran2024,Chen2025,Lobentanzer2025}, triggering a surge in publications across a variety of disciplines like Medicine, Neuroscience, and Molecular Biology \citep{Schmallenbach2024}. State-of-the-art models like Google Gemini, OpenAI GPT, and DeepSeek are applied to a wide range of tasks, including but not limited to biomedical literature and knowledge mining (e.g., information extraction from electronic health records \citep{Yang2022}), drug discovery and genomics (e.g., deciphering protein structures \citep{Valentini2023}), and clinical diagnoses and decision support \citep{Zhou2025,Chen2025}. This rapid integration signals a fundamental shift in the toolkit available to the biomedical community.

Beyond their technical breakthroughs, LLMs are fundamentally reshaping how biomedical knowledge is produced, validated, and communicated \citep{Koller2024,Zhang2025,Meng2024,Li2025_TrBiotech}. Their capacity to synthesize knowledge across highly specialized subdomains \citep{Li2025_TrBiotech} makes them pivotal for fostering interdisciplinary collaboration, where groundbreaking discoveries are emerging extensively at the intersection of multiple fields. By translating complex concepts into more accessible language, LLMs help biomedical researchers from diverse backgrounds communicate more effectively, overcoming challenges created by technical barriers or methodological differences \citep{Moon2024}. Meanwhile, LLMs can lower the entry bar for researchers to AI studies by assisting with tasks such as coding and data analysis \citep{Moon2024}. Therefore, LLMs may play a dual role: they promote cross-disciplinary collaboration; simultaneously, they empower non-expert researchers to independently develop and deploy sophisticated AI-based solutions, lowering computational barriers in biomedical research \citep{Moon2024}.

This transformative promise of LLMs, however, can be shadowed by inequitable resource distribution and access. The effective deployment of LLMs often requires substantial computational resources, technical infrastructure, and institutional support \citep{Ahmed2023,Lauer2021}. These prerequisites disproportionately favor well-resourced institutions \citep{Lauer2021}, such as top-ranked universities and major academic medical centers, where high-performance computing environments and interdisciplinary teams are readily available. As a result, LLM-related biomedical research risks becoming highly concentrated within well-resourced organizations, leaving smaller laboratories and under-resourced institutions at a disadvantage \citep{Kudiabor2024}. Such concentration could further exacerbate existing disparities in research capacity, visibility, and impact in the scientific community.

As such, significant gaps remain in understanding the current landscape of LLM-related biomedical research—specifically, who participates, how collaborations arise, and where capacity gaps persist. Previous studies have largely examined LLMs’ effects on scientific outputs through bibliometric analysis and literature reviews, with some leveraging co-authorship networks from sources like OpenAlex and Scopus to identify leading institutions or countries driving LLM-related biomedical research \citep{Gencer2025,Yu2024,Carchiolo2025,Liu2024}. Yet these studies have not explicitly investigated fundamental questions including how LLM research is reshaping collaboration in the biomedical field, the temporal trends of collaboration diversity across institutions, disciplines, and countries, and relationships between collaboration structure, research output/impact, and institutional resources. Our study addresses these gaps by investigating three main questions: (1) the emerging inter-institutional and interdisciplinary collaboration pattern in LLM-related biomedical research; (2) which institutions and disciplines serve as central and bridging entities within the collaboration networks; and (3) how the institutional participation patterns reflect disparities between well-sourced and under-resourced organizations.

We address these gaps through a large-scale analysis of PubMed-indexed LLM-related biomedical articles between October 1, 2018, and February 7, 2025 ($n=5{,}674$), along with two control groups, machine learning (ML) and general biomedical research (sampled from the same recent publication period; $n=100{,}000$ each). Our findings show that, within the LLM-related biomedical research domain, while a few institutions remain dominant in central positions in collaboration networks, LLMs are lowering technical barriers and inspiring revolutions, facilitating broader participation from disciplines outside Computer Science (CS) and Artificial Intelligence (AI) in key biomedical research advancements. Importantly, our results also reveal that partnering with central hubs might offer resource-constrained institutions a pathway to greater visibility and impact. Overall, this study provides insights into the evolving patterns of collaboration, resource distribution, and the mechanisms shaping equitable growth in LLM-related biomedical research, thereby contributing to the emerging “science of science” \citep{Fortunato2018} in biomedicine.

\section{Methods}

\subsection{Data preparation}
We selected PubMed as our data source given that it is the most comprehensive and authoritative database for biomedical and healthcare research, containing more than 35 million citations from MEDLINE, life science journals, and online books \citep{Lu2011}. Our data preparation began by collecting LLM-related papers indexed in PubMed between October 1, 2018 (approximately when Google’s BERT \citep{Devlin2019} was released) and February 7, 2025. All downloads, filtering, and tabulations were performed using NCBI E-utilities \citep{Sayers2025}. We searched for relevant papers using either general terms (e.g., “large language model”, “LLM”) or specific model names included in the MMLU benchmark \citep{Hendrycks2021}. This search covered both open-source models (e.g., BERT, Flan-T5, LLaMA) and closed-source models (e.g., GPT, Claude). To improve data quality, we excluded models that had no PubMed records (e.g., Galactic) or produced excessive noise (e.g., Yi, Phi). To focus on original research and further ensure relevance, we implemented a three-stage data cleaning process. First, we restricted the dataset to journal articles and preprints, excluding reviews, comments, editorials, opinion pieces, and errata, yielding 6,412 LLM-related records. Second, we removed duplicates by comparing titles and using Jaccard similarity to identify papers published both as preprints and final articles. Third, we used the GPT-4-as-a-judge model \citep{Zheng2023,Gu2025} to evaluate the relevance of each paper based on its title, abstract, and keywords. The final cohort included 5,674 LLM-related papers.

To assess whether the identified collaboration patterns are unique to LLM research or reflect the trend of natural progression of a general field, we established two control groups. The first control group consisted of ML research papers, given that ML is a well-established field from which LLM emerged as a subfield. This group contained 98,035 randomly sampled papers containing the term “machine learning” in the title or abstract. To further compare beyond AI-focused fields, we established a second control group of 96,313 papers randomly selected from the general biomedical research that is unrelated to ML or LLM, allowing comparison of collaboration patterns across a wider scientific landscape. Finally, to ensure consistency in institutional representation, we standardized institution names across all datasets using GPT-4o. We applied this harmonization to 200,363 papers across all three datasets (LLM, ML, and general biomedical research) to ensure consistency and accuracy in downstream analyses.

\subsection{Measure of collaboration diversity: Shannon entropy}
To quantify collaboration diversity, we applied Shannon entropy \citep{Shannon1948}, a measure that captures uncertainty in a distribution. Let $x_i$ denote the $i$-th author in a publication and $y_j(x_i)$ the attribute associated with that author (e.g., department, institution, or country). For a given publication with $k$ authors $(x_1, x_2, \dots, x_k)$ and $n$ distinct affiliation attributes $(y_1, y_2, \dots, y_n)$, let $P(Y)=(p(y_1), p(y_2), ..., p(y_n))$ be the empirical probability distribution of the attribute variable $Y$, where $p(y_j)$ is the proportion of authors affiliated with attribute $y_j$. Then the entropy for that paper is:
\[
H(Y) = -\sum_{j=1}^{n} P(y_j)\log P(y_j).
\]
In this study, we examined collaboration diversity across three dimensions of author affiliation: institution $I(x_i)$, department $D(x_i)$, and country $C(x_i)$, denoted by $Y(x_i) = \{I(x_i), D(x_i), C(x_i)\}$. 

As an illustrative example, consider a publication with $k = 5$ authors affiliated with $n = 3$ different departments: $\{d_1(x_1), d_1(x_2), d_2(x_3), d_2(x_4), d_3(x_5)\}$. The three departments, $d_1$, $d_2$, and $d_3$, have probabilities $0.4$, $0.4$, and $0.2$, respectively. 

The Shannon entropy for department is therefore:
\[
H(D) = - \big(0.4 \log_2 0.4 + 0.4 \log_2 0.4 + 0.2 \log_2 0.2\big) = 1.522
\]

We computed Shannon entropy separately for each attribute (institution, discipline, and country). Higher entropy values reflect greater diversity in collaboration, whereas lower values indicate more uniform affiliations. An entropy of $0$ indicates all authors sharing the same affiliation.

\subsection{Measure of collaboration structure: network theory}
To characterize collaboration structure, we applied network analysis techniques to reveal both structural relationships and key influential actors among research communities. We considered two types of nodes: institutions and disciplines. We used two centrality measures: degree centrality, reflecting a node’s prominence based on the number of its direct connections, and betweenness centrality, which identifies nodes that serve as critical bridges between communities. Together, these metrics helped us identify influential entities (institutions or disciplines) that facilitate cross-institutional and cross-disciplinary collaboration in LLM-related biomedical research.

For network visualization and community detection, we implemented a two-step approach. First, we applied the Clauset-Newman-Moore (CNM) algorithm \citep{Clauset2004} to detect densely connected research clusters, thereby revealing the community structures within the collaboration network. Second, we utilized the Fruchterman-Reingold force-directed layout \citep{Fruchterman1991} to generate visual representations of the collaboration networks. This layout algorithm optimized node positioning by minimizing edge crossings and ensuring even distribution of the nodes, yielding interpretable, tree-shaped visualizations. We further evaluated the global network properties, including vertices (researchers’ affiliated entities), edges (collaborations between entities), shortest distance (the smallest number of collaboration links required to connect a pair of entities), average path length (the average shortest distance between all pairs of entities of the network), number of communities (densely connected groups of entities), number of components (subgraphs in which every entity is reachable from any other within the same subgraph, with no connections across subgraphs), modularity (the degree to which the network can be divided into well-defined communities), edge density (the ratio of observed edges to all possible edges), coverage (fraction of edges that fall within communities), and community size (number of entities in each community).

\subsection{Measure of institutional resources}
We used the 2024 fiscal year (FY) National Institutes of Health (NIH) funding data, downloaded from the Research Portfolio Online Reporting Tools (RePORT) \citep{NIHRePORT}, as a proxy for institutional biomedical research resources at US-based institutions. This is consistent with prior work that leverages NIH support to indicate research capacity and productivity \citep{Hendrix2008}. While LLM-related biomedical publications span multiple years, we used the FY 2024 (Oct 1, 2023 – Sep 30, 2024) data, assuming that institutional resources are relatively stable in recent years and are reflective of the ongoing capacity to engage in LLM research. We further assessed the association between resources and institutional performance using three key indicators derived from our PubMed article corpus: (1) degree centrality (collaboration), (2) publication volume (research output), and (3) citations per paper (research impact). These metrics provide a multidimensional view of institutional-level engagement in LLM-related biomedical research.

Due to the limited availability of comprehensive global funding data and the challenges of cross-country comparisons arising from different institutional funding mechanisms, we restricted our analysis to U.S. institutions. In addition, as institution names often vary between NIH records and PubMed datasets, we developed a harmonization pipeline integrating semantic embeddings (BAAI/bge-m334 \citep{Chen2024_BGE})) with LLM-based verification (GPT-4o-mini) to ensure precise alignment of institution names across data sources. After this alignment, we examined associations between resources and the three institutional participation metrics using Spearman correlation coefficients. We then modeled and visualized these relationships in log–log space with a polynomial regression fit. Finally, we identified institutions with performance differing from expectations, those with residuals more than 1.5 SD above or below the model prediction.

We further drew our attention to institutions with below-median resources and mapped these institutions onto our previously constructed collaboration network. Within this stratum, we classified institutions into two groups: ``high achievers,'' whose citations per paper were higher than expected given their resources, and a comparison group comprising the remaining institutions. We then tested whether high achievers allocated a larger share of their collaborations to network hubs or bridges. We defined hubs and bridges as institutions in the top 1\% ($n=59$), 2.5\% ($n=146$), or 5\% ($n=291$) by degree and betweenness centrality, respectively; varying these cutoffs served as a sensitivity analysis for the hub/bridge definition. For each institution, we computed the share of collaborations involving hubs (or bridges) by dividing its number of edges to hubs (or bridges) by its total number of edges. Before comparing groups, we attenuated the influence of extreme values within each group by trimming observations that fell outside $[Q1 - 1.5\,\mathrm{IQR},\, Q3 + 1.5\,\mathrm{IQR}]$. We then conducted one-sided Wilcoxon rank-sum tests, with the alternative that the high achiever distribution was shifted to the right of the comparison group.

\section{Results}

\subsection{LLM research facilitates broad scientific collaboration across institutions and disciplines while lowering the barrier for non-CS researchers to make an impact}

\begin{figure}[t]
  \centering
  \includegraphics[width=\linewidth]{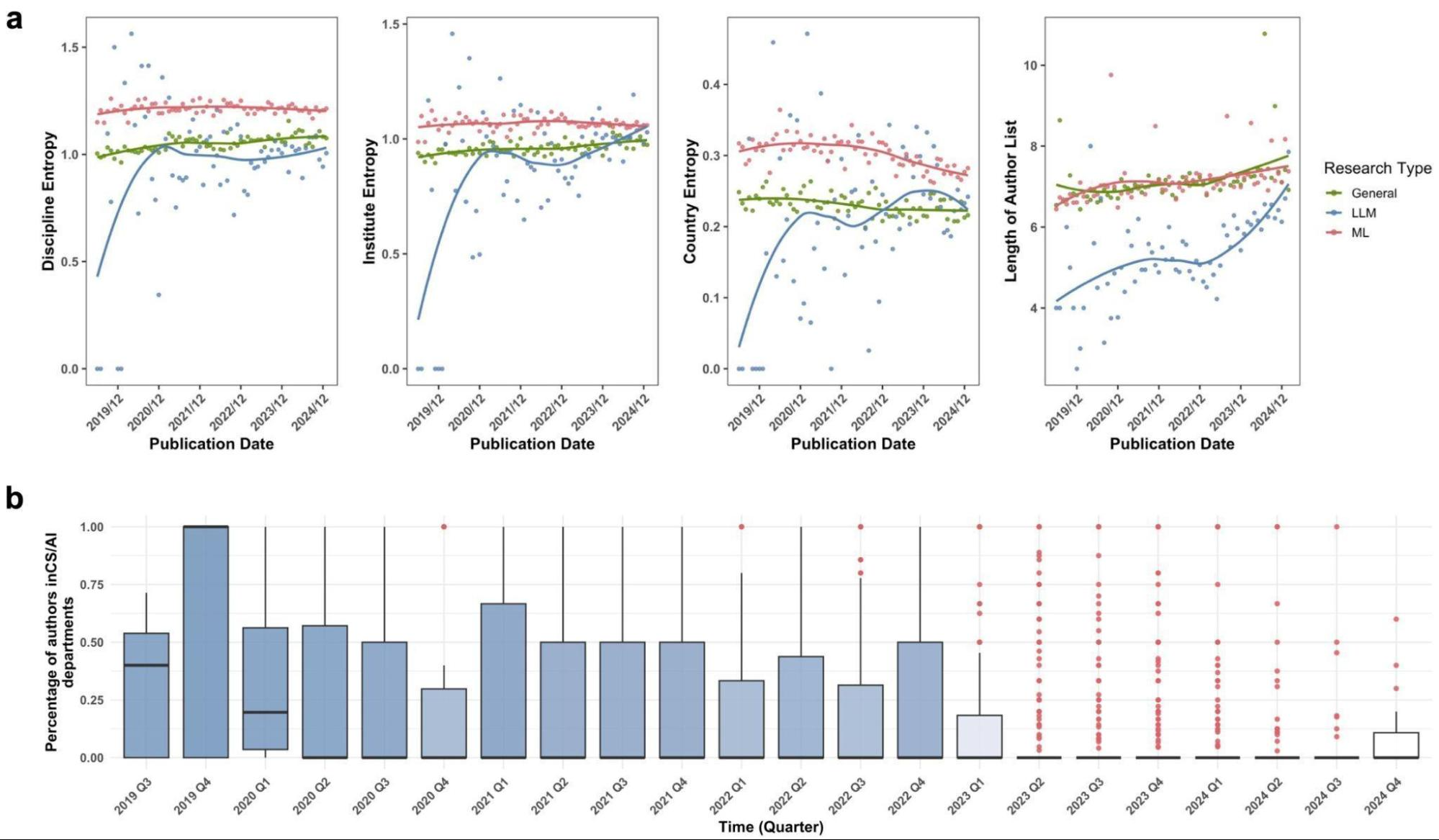}
  \caption{Collaboration patterns and CS/AI participation in LLM, ML, and general research in biomedicine based on publications recorded in PubMed between June 2019 and January 2025. (a) Change of discipline, institution, and country entropy over time in LLM, ML, and general biomedical publications. (b) Change in the proportion of CS/AI authors in LLM publications.}
  \label{fig:entropy}
  \phantomsubcaption\label{fig:entropy:a}
  \phantomsubcaption\label{fig:entropy:b}
\end{figure}

We investigated biomedical publications in three fields: (1) LLM-related, (2) ML-related excluding LLM, and (3) general research excluding ML or LLM-related, referred to as general biomedical research. We first examined the temporal changes in \emph{discipline entropy}, \emph{institution entropy}, and \emph{country entropy} for publications in these three domains between June 2019 (the month the first PubMed-indexed LLM publication appeared) and January 2025 (\textbf{\hyperref[fig:entropy]{Figure~\ref{fig:entropy}a}}). Overall, the discipline and institution entropies in ML-related research in biomedicine have stabilized over time, while the discipline and institution entropies in general biomedical research show a stable and slightly increasing pattern, which is consistent with the maturity and relative stability of collaboration structures in both fields in recent years. The country entropy in the two fields, especially the ML research field, shows a decreasing trend over the study period, suggesting that international collaborations in these fields have become more geographically concentrated. This trend may be partially influenced by policies governing data security, funding, and knowledge exchange that have increasingly shaped the landscape of global science. Compared to the general research in biomedicine, ML research shows higher discipline, institution, and country entropy, reflecting its interdisciplinary nature. In comparison, LLM shows more variable entropy patterns relative to ML and general biomedical research, reflecting its status as a rapidly evolving field that has not yet stabilized. For example, the average discipline entropy in LLM research has been fluctuating around 1.0 and shows a potentially increasing trend, though still slightly below that observed in general biomedical research. The average institution entropy in LLM research initially increased, then decreased, before increasing again to reach a level comparable to general ML research. In contrast, the average country entropy in LLM research peaked at the end of 2023 at a level higher than that in general biomedical research, before showing a clear decreasing trend afterwards.

We next examined the change in the participation of CS and AI disciplines in LLM-related research since the first emergence of LLM (\textbf{\hyperref[fig:entropy]{Figure~\ref{fig:entropy}b}}). Specifically, we tracked the proportion of authors from CS/AI-affiliated disciplines over time. We observed that the mean proportion of CS/AI authors decreased from a high level of 50\%–100\% during the early stage in 2019 to nearly zero in the last two years. This observation, along with the trend of discipline entropy changes shown in \textbf{\hyperref[fig:entropy]{Figure~\ref{fig:entropy}a}}, suggests that LLM research was initially driven almost exclusively by CS/AI investigators, then expanded through interdisciplinary collaboration between CS/AI and other domains, and has more recently been led or co-led by researchers outside the CS/AI fields. In other words, the bar for conducting LLM research has been lowered for researchers outside of the CS/AI fields over the last five years. As LLM algorithms and technologies continue to be refined, the CS/AI expertise that was highly demanded during the very early stage has diffused into other domains. Nevertheless, we continue to observe publications coming out that were produced entirely by CS/AI researchers every quarter, which suggests the ongoing efforts on revolutionizing LLM technologies led by CS/AI researchers (March 2025).

We further examined how collaboration diversity in disciplines, institutions, and countries affects the impact of an LLM research paper. Specifically, we conducted regression analyses to check the association between the impact of LLM research papers, reflected by Relative Citation Ratio (RCR) \citep{Hutchins2016}, citations per year, and expected citations per year (all were log-transformed to be approximately normally distributed), and discipline, institution, and country entropy, adjusting for covariates including publication date (by month) and the proportion of CS authors. Here, RCR is a metric of scientific influence that represents the field- and time-normalized citation rate, which is benchmarked to 1.0 for a typical (median) NIH-funded paper in the corresponding year of publication. For LLM research, none of the three entropy measures is significantly associated with RCR, citations per year, or expected citations per year ($p>0.05$), suggesting that, even with less domain-specific knowledge and cross-disciplinary collaborations, an LLM-related research paper could still yield high impact. On the contrary, for ML and general biomedical research, higher discipline entropy and country entropy are consistently and significantly associated with higher RCR, higher citations per year, and higher expected citations per year ($p<0.001$ to $p=0.001$), while higher institution entropy is significantly associated with higher RCR and citations per year ($p$ from $2\times10^{-16}$ to $5.7\times10^{-7}$) and not significantly with expected citations per year ($p>0.05$). These findings suggest that LLM-related biomedical research publications with lower collaboration entropy could achieve the same level of impact as those produced by high-level cross-disciplinary, institutional, and country collaborations. A possible explanation for such a pattern is that LLM research has lowered the barrier for scientists in biomedical fields who wish to adopt state-of-the-art LLM within their disciplines without high demand for external support, especially from CS collaborators. Overall, we observed that, while cross-disciplinary and cross-institution biomedical collaboration are benefiting from LLM research topics with increasing collaboration diversity, the continuously emerging LLM tools and inspired topics are also facilitating non-CS researchers to make an impact in the field more independently.

\subsection{Well-established institutions and leading disciplines such as Medicine and CS anchor the collaborations}

While LLM-related biomedical research has become increasingly accessible across institutions and disciplines, such inclusivity may still be shaped mainly by the structural roles played by some leading entities within the collaboration ecosystem. It is thus important to examine the key actors, including institutions and disciplines, that serve as central hubs or critical bridges to anchor and connect research networks. Such investigation can help explain how ideas and resources flow across the community and set the stage for understanding how these structural roles relate to broader resource disparities.

\begin{figure}[!htbp]
  \centering
  \includegraphics[width=1\linewidth]{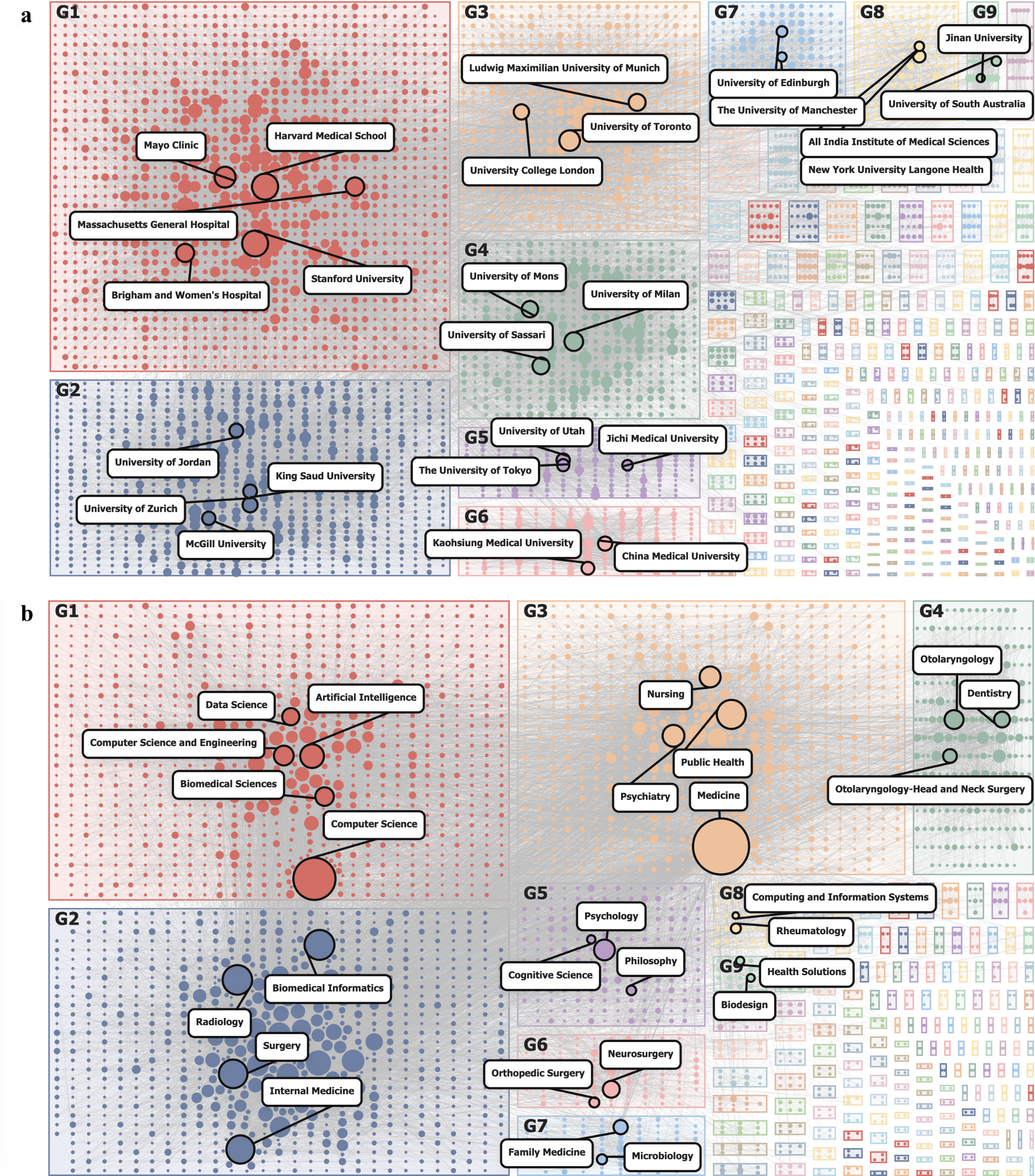}
  \caption{Co-authorship networks based on authors’ (a) institution and (b) disciplinary affiliations. Each node represents a research institution or discipline, and each edge represents a co-authorship between researchers from the two respective institutions or disciplines. Node colors match the colors of the clusters (G1–G9) to which the nodes belong, where the clusters were determined based on the CNM algorithm. The network layout represents the community clusters determined based on the Fruchterman–Reingold force-directed layout algorithm.}
  \label{fig:networks}
  \phantomsubcaption\label{fig:networks:a}
  \phantomsubcaption\label{fig:networks:b}
\end{figure}

The co-authorship networks presented in \textbf{\hyperref[fig:networks]{Figure~\ref{fig:networks}a}} and \textbf{\hyperref[fig:networks]{Figure~\ref{fig:networks}b}} show the institutional and interdisciplinary collaboration patterns in LLM-related biomedical research, respectively. \textbf{\hyperref[tab:metrics]{Table~\ref{tab:metrics}}} summarizes the network metrics characterizing both institutional and interdisciplinary collaboration networks. In the institutional network, several prominent institutions emerged as central nodes, including Stanford University and Harvard Medical School in the largest collaboration cluster G1, the University of Zurich in the second largest cluster G2, and University College London in the third largest cluster G3. The network consisted of 5,804 institutions connected through 24,544 collaborative relationships, forming 301 distinct communities. Stanford University showed the highest degree centrality (297 connections), followed by Harvard Medical School (266) and the Mayo Clinic (194). Regarding bridging roles, Stanford University still shows the highest betweenness centrality (0.0758), followed by Harvard Medical School (0.0629) and the Mayo Clinic (0.0518), highlighting their pivotal roles in facilitating cross-institutional collaboration.

\textbf{\hyperref[fig:networks]{Figure~\ref{fig:networks}b}} illustrates the discipline-level collaboration network, reflected by co-authorship across different institutional departments, in biomedical LLM research. The visualization reveals distinct clusters of research disciplines in collaboration, with key areas including Public Health and Surgery in G1, CS and AI in G2, and Biomedical Informatics and Radiology in G3. The interdisciplinary collaboration network encompassed 3,706 disciplines connected through 17,451 collaborative links, forming 179 distinct communities. Medicine emerged as the discipline that has the highest degree centrality (836 connections), followed by CS (609), Public Health (328), Surgery (289), and Internal Medicine (283). Regarding the betweenness centrality, Medicine (0.1849) and CS (0.1449) showed the highest values, suggesting their crucial role in connecting different research domains.

\begin{table*}[!htbp]
\centering
\caption{Network metrics based on authors’ institutional and disciplinary affiliations.}
\label{tab:metrics}
\renewcommand{\arraystretch}{1.1}
\begin{tabularx}{\textwidth}{lYY}
\toprule
\textbf{Metric} & \textbf{Institutional Affiliation (CNM)} & \textbf{Disciplinary Affiliation (CNM)} \\
\midrule
Num. communities & 301 & 179 \\
Num. connected components & 225 & 126 \\
Num. edges & 24{,}544 & 17{,}451 \\
Num. vertices & 5{,}804 & 3{,}706 \\
Modularity & 0.6548 & 0.4243 \\
Avg. clustering & 0.6809 & 0.7061 \\
Avg. path length & 4.1731 & 3.1042 \\
Edge density & 0.0015 & 0.0025 \\
Coverage & 0.8276 & 0.6444 \\
Top 10 community sizes & 1445, 973, 661, 656, 340, 163, 111, 94, 62, 54 & 849, 724, 644, 314, 291, 141, 44, 38, 34, 27 \\
Top 10 degree entities & 
Stanford (297), Harvard Medical School (266), Mayo Clinic (194), University of Toronto (167), Massachusetts General Hospital (149), University of Washington (148), University of Oxford (142), Ludwig Maximilian University of Munich (132), Johns Hopkins University (129), Brigham and Women’s Hospital (127)
&
Medicine (836), Computer Science (609), Public Health (328), Surgery (289), Internal Medicine (283), Biomedical Informatics (280), Radiology (271), Neurology (264), Artificial Intelligence (224), Pediatrics (200) \\
Top 10 betweenness entities &
Stanford University (0.0758), Harvard Medical School (0.0629), Mayo Clinic (0.0518), University of Toronto (0.0365), University of Oxford (0.0305), University of Washington (0.0297), Johns Hopkins University (0.0271), Ludwig Maximilian University of Munich (0.0268), Shanghai Jiao Tong University (0.0241), Tsinghua University (0.0236)
&
Medicine (0.1849), Computer Science (0.1449), Public Health (0.0466), Surgery (0.0348), Radiology (0.0346), Internal Medicine (0.0341),  Psychology (0.0328), Artificial Intelligence (0.0301), Neurology (0.0291), Biomedical Informatics (0.0289) \\
\bottomrule
\end{tabularx}
\end{table*}

The network metrics presented in \textbf{\hyperref[tab:metrics]{Table~\ref{tab:metrics}}} highlight several key characteristics for the two networks. For the institutional collaboration network, the low edge density (0.0012) coupled with the high average clustering coefficient (0.6902) suggests that institutions tend to form tight-knit research clusters rather than dispersed connections. The average path length (4.2771) indicates that research knowledge typically traverses multiple steps to flow between institutions. The network coverage (0.8318) shows that a substantial majority of institutions are connected to the main network components, which are mainly composed of large, densely connected communities: top four communities have 1,445, 973, 661, and 656 members, respectively, accounting for 64\% of the institutions. In the discipline-level collaboration network, similar patterns emerged but with notable distinctions. The edge density (0.0025) and clustering coefficient (0.7061) indicate comparable clustering tendencies to institutional collaboration. However, the shorter average path length (3.1042) suggests that connections between disciplines are more direct than those between institutions. The discipline community size distribution also follows a hierarchical structure, but with smaller leading communities (849, 724, 644, and 314 members in the four largest communities), indicating a more decentralized collaboration pattern across different medical and technical disciplines.

\subsection{Institutional research output is driven by resource levels and further promoted via strategic collaboration and network positioning}

The prominence of a relatively small set of central and bridging institutions within collaboration networks raises questions about how research resources underpin these positions of influence. LLM-related research, like other computationally intensive domains, demands substantial infrastructure, funding, and technical resources. Therefore, the centrality patterns we observed in the collaboration network may partly reflect underlying disparities in institutional resources. We therefore examined whether institutional resource levels, focusing on U.S. institutions and using NIH funding in FY 2024 as a proxy, correlate with key performance metrics: research output (total publications during the study period), collaboration centrality (degree in the institutional network), and research impact (citations per paper).

\begin{figure}[!htbp]
  \centering
  \includegraphics[
  width=\linewidth,
  trim=4mm 4mm 4mm 4mm, 
  clip
  ]{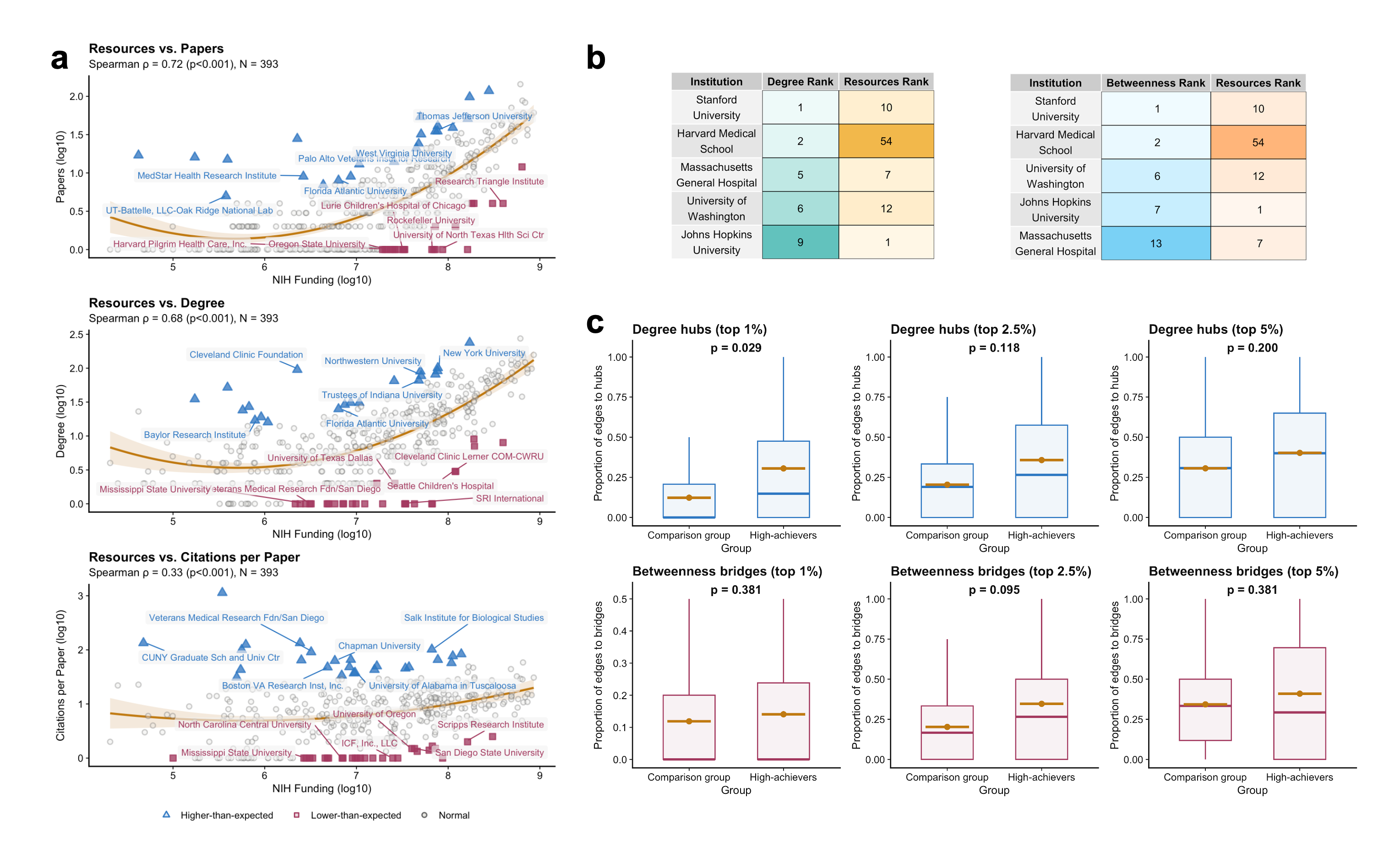}
  \caption{Institutional research output is influenced by institutional resource levels as well as collaboration and network positioning. (a) Polynomial fits (log–log scale) relating FY2024 NIH funding (biomedical research resources) to LLM-related outcomes: degree centrality, publication count, and citations per paper. Shaded bands are 95\% confidence intervals of the fitted curves (highlighted in yellow). Outliers are institutions with residuals $> \pm 1.5$ SD from the fitted trend. Points above the funding-based expectation are shown in blue; those below are in red. (b) Top five U.S. institutions by degree or betweenness rank with unambiguous NIH awardee-name mapping. Columns report each institution’s degree/betweenness rank and its FY2024 NIH funding rank. (c) Proportion of collaborations with network hubs (and, separately, with bridges), for institutions in the below-median NIH-funding group, contrasting ‘high achievers’ (higher-than-expected citations per paper) with the remainder. Hubs/bridges are defined using degree/betweenness centrality thresholds (top 1\%, 2.5\%, or 5\%). Boxplots show distributions with group means overlaid; one-sided Wilcoxon rank-sum tests evaluate whether high achievers allocate a larger share to hubs.}
  \label{fig:resources}
  \phantomsubcaption\label{fig:resources:a}
  \phantomsubcaption\label{fig:resources:b}
  \phantomsubcaption\label{fig:resources:c}
\end{figure}

Our analytic sample included only U.S. institutions whose names could be mapped accurately between NIH records and our PubMed authorship data, yielding a total of 393 institutions. We observed strong positive associations on the log–log scale between institutional resources and each performance metric (\textbf{\hyperref[fig:resources]{Figure~\ref{fig:resources}a}}), with Spearman correlations $\rho=0.72$ ($p<0.001$) for total number of publications, $\rho=0.68$ ($p<0.001$) for collaboration degree, and $\rho=0.33$ ($p<0.001$) for citations. This pattern suggests that greater resources are more closely linked to producing more papers and occupying more central positions in collaboration networks than to achieving higher citation impact. Fitted polynomial regression curves with 95\% confidence intervals further illustrate nonlinear trends across the three metrics, with steeper increases at higher resource levels.

As shown in \textbf{\hyperref[fig:resources]{Figure~\ref{fig:resources}b}}, institutions that served as hubs in the collaboration network (high degree centrality) and bridged its communities (high betweenness centrality), including the University of Washington, Harvard Medical School, Massachusetts General Hospital, Johns Hopkins University, and Stanford University, also ranked among the top holders of biomedical research resources, suggesting these resources could have helped underpin their influence. However, the relationship between institutional resources and research performance was far from deterministic. \textbf{\hyperref[fig:resources]{Figure~\ref{fig:resources}a}} shows that several institutions deviate substantially from the expected trend, with red and blue points indicating those more than 1.5 standard deviations above or below their expected performance given resource levels. This pattern held across all three performance dimensions: number of papers, degree centrality, and citations per paper. Some institutions (e.g., MedStar Health Research Institute and Florida Atlantic University) outperformed their biomedical research resources in research output, while others (e.g., Baylor Scott \& White Research Institute and the Cleveland Clinic Foundation) showed higher degree centrality, possibly reflecting their “prioritization” of LLM-related biomedical research. By contrast, hospitals such as Lurie Children’s Hospital of Chicago underperformed on research output, and Seattle Children’s Hospital underperformed on degree centrality, perhaps because they allocated a larger share of resources to translational research and basic science rather than state-of-the-art AI research.

Visual inspection of the boxplots (\textbf{\hyperref[fig:resources]{Figure~\ref{fig:resources}c}}), with group means overlaid, suggests that high achievers, whose citations per paper were higher than expected given their resources, generally have equal or higher proportions than the comparison group, but the strength of evidence varies by threshold and by hub/bridge. For degree-based hubs, the difference is statistically significant at the top 1\% threshold (one-sided Wilcoxon rank-sum $p=0.029$), while results at 2.5\% ($p=0.118$) and 5\% ($p=0.200$) are not significant. For betweenness-based bridges, contrasts are not significant at 1\% ($p=0.381$) and 5\% ($p=0.381$), with a suggestive but non-significant trend at 2.5\% ($p=0.095$). Our findings suggest that overachieving institutions compared to their resource levels tend to allocate a larger share of their collaborations to most influential hub institutions, defined as those with the highest degrees in our network, and this tendency is most pronounced for collaborations with the top 1\% of hubs.

\section{Discussion}
This study investigated the dynamics and structures of scientific collaboration in LLM-related biomedical research through a large-scale analysis of 5,674 PubMed-indexed articles. Our analysis revealed a declining share of CS and AI authors in LLM-related studies over time, reflecting the increasing accessibility of LLM research to general investigators in biomedical fields. Unlike the relatively stable collaboration patterns seen in general ML and biomedical research, team diversity in LLM-related studies continued to expand, suggesting that the field’s collaborative landscape is still evolving. At the paper level, we did not find a significant association between the diversity entropy of author teams and citation impact. This absence of correlation, however, should not be interpreted as evidence that collaboration is unimportant in biomedical LLM research. Rather, it may reflect the unique nature of this domain that LLMs may lower technical barriers, enabling smaller or less diverse teams to achieve high-impact outcomes. Network analyses further identified leading institutions, such as Stanford University and Harvard Medical School, as hub or bridging institutions, with disciplines like Medicine, Computer Science, and Public Health serving as key bridging disciplines. At the institution level, the performance metrics remain strongly correlated with resources. However, our results indicate that collaboration strategy matters for under-resourced institutions: those with more collaborations with central hubs tend to achieve greater visibility and impact in the LLM-related biomedical research area.

LLMs appear to drive broader and more diverse collaboration by lowering technical barriers to entry: we observed a clear trend of increasing collaboration diversity across institutions and disciplines in LLM-related biomedical research, as measured by Shannon entropy \citep{Shannon1948}, which exceeded that in both broader ML and general biomedical fields. This was further supported by a significant decline over time in the proportion of authors from CS and AI disciplines on LLM-related papers. These patterns suggest that LLM-assisted biomedical research enables broader participation from non-technical domains. As LLMs become more accessible and user-friendly, researchers in areas such as Medicine, Public Health, and Internal Medicine are increasingly able to leverage their expertise for diverse tasks in LLM relevant studies without the need for formal collaboration with CS/AI specialists. This observation also indicates that LLM-based studies are becoming a shared ground, helping to bridge methodological and communication gaps across disciplinary boundaries. In addition, for LLM-related papers, entropy-based diversity measures across disciplines, institutions, and countries showed only moderate associations with citations, unlike in broader ML and general biomedicine, where the association was stronger. A likely explanation is that LLMs are lowering technical barriers, allowing smaller or less diverse teams to achieve similarly high impact.

Collaboration networks in LLM-related biomedical research are anchored by hub institutions and key bridging disciplines: Although collaboration diversity across institutions and disciplines has increased, our network analysis indicated that the research ecosystem remained highly centralized, with activity concentrated within a core group of influential institutions and disciplines. At the institution level, prominent hubs such as Stanford University, Harvard Medical School, and the Mayo Clinic dominated the network, shaping collaboration patterns and knowledge flows. However, the relatively long average path length of the institutional network suggested fewer collaborative pathways independent of hub institutions. At the discipline level, Medicine and CS emerged as critical nodes, linking a diverse array of disciplines—including public health, radiology, and biomedical informatics—and facilitating the integration of AI methodologies into clinical research. High clustering coefficients and modularity values indicated efficient collaboration and knowledge exchange within tightly connected communities. Notably, bridging disciplines such as biomedical informatics appeared to play a central role in translating foundational AI advances into diverse clinical domains, including Surgery, Radiology, and Internal Medicine, thereby serving as a key bridge between technical innovation and clinical application.

Institutional resources strongly influence research performance, but strategic collaboration could partially offset resource limitations: Our analysis revealed consistently positive associations between institutional biomedical research resources (FY2024 NIH funding) and key performance metrics, including collaboration centrality, research output, and research impact, with fitted curves rising steeply at higher resource levels. Nevertheless, several institutions exceeded expectations relative to their resources, attaining high centrality, publication volume, and citations. Within the below-median resource group, high achievers (higher-than-expected citations per paper) were characterized by stronger ties to hubs in the collaboration network, i.e., organizations with high degree centrality, which were partners that anchored the biomedical LLM field through numerous co-authorships, especially when ‘hubs’ were defined narrowly (top 1\%). Such links plausibly expanded reach, accelerated idea diffusion, and increased visibility, any of which could help translate constrained resources into greater scientific impact. In short, while observational, our results suggest that resource-constrained institutions that collaborated with the most highly connected partners tend to perform better in biomedical LLM research. These findings should not be interpreted causally; they may reflect selection, visibility, or topical alignment rather than a direct payoff from partnering with central actors in the collaboration network.

Our study presents several opportunities for future research. First, our reliance on PubMed as the primary data source is an inherent limitation. Although PubMed is the most comprehensive repository for biomedical and life sciences literature, it does not fully capture relevant CS research, particularly the work published in conference proceedings or on preprint servers such as arXiv, which are not indexed in PubMed but are often central to advances in CS and AI fields. Second, while our LLM-based pipelines for relevance screening and affiliation matching can improve data quality, these automated processes are subject to potential errors, especially given the heterogeneity of institutional naming conventions across PubMed and NIH records. Third, future research could use a more comprehensive proxy for institutional resources for LLM biomedical research instead of total NIH funding. Finally, given the rapid pace of innovation in the field and continual emergence of new foundational models such as GPT-5, continuous data collection and longitudinal analyses could be essential to capture the evolving landscape of LLM-driven research and its broader impact on the biomedical research community.

Overall, our findings demonstrate that collaboration continues to play a central role in this rapidly evolving area. Our study helps the scientific community better understand who is participating in and benefiting from LLM-driven biomedical advances. Our study also highlights persistent disparities in institutional participation and underscores the need for strategies that broaden access to expertise, data, and computational resources to enable more equitable engagement with LLM-driven advances.

\section*{Code and Data Availability}
The code for reproducing the results in this study is available at \url{https://github.com/syma-research/LLM_PubMed_SoS/}. The data used in our analysis can be downloaded at \url{https://drive.google.com/drive/u/0/folders/1WhFyzpVeIWHFCbChCx3LCM5OI7AxqWdB}.

\section*{Acknowledgements}
This work was supported by the following National Institutes of Health (NIH) grants: R00~HG012223 (J.J.) and R35GM157133 (J.J.).

\section*{Competing interests}
All authors declare no competing financial interests.

\bibliographystyle{unsrtnat}
\bibliography{references}

\end{document}